\documentclass[journal]{IEEEtran}

\usepackage{cite}
\usepackage{amsmath,amssymb,amsfonts}
\usepackage{textcomp}
\usepackage{xcolor}
\usepackage{graphicx}
\usepackage{booktabs}
\usepackage[hyphens]{url}
\usepackage{ragged2e}
\usepackage{multirow}

\begin{document}

\title{Direction-Preserving Active Noise Control with a Conditional Control-Filter Estimation Network}

\author{Ziyi~Yang,
        Zhengding~Luo,
        Boxiang~Wang,
        Libin~Zhang,
        and~Woon-Seng~Gan%
\thanks{This work was supported by the Ministry of Education, Singapore, through Academic Research Fund Tier 2 under Grant MOE-T2EP20224-0010.\par Ziyi Yang, Zhengding Luo, Boxiang Wang, and Woon-Seng Gan are with the Smart Nation TRANS Lab, School of Electrical and Electronic Engineering, Nanyang Technological University, Singapore (email: \{ziyi016, luoz0021, boxiang001\}@e.ntu.edu.sg; ewsgan@ntu.edu.sg). Zhengding Luo is the corresponding author.\par Audio demo: \protect\url{https://yzyzieee.github.io/DP-ANC-Demo/}. Source code will be released at \protect\url{https://github.com/yzyzieee/DP-ANC}.}%
}

\markboth{}{}

\maketitle

\begin{abstract}
\boldmath
Conventional active noise control (ANC) minimizes the total disturbance at the error microphone without distinguishing desired sound from noise. Direction-preserving ANC (DP-ANC) instead aims to attenuate a noise component arriving from a direction other than the specified desired direction while preserving sound naturally arriving from that direction. Existing approaches typically either require analytical optimization to be repeated for each new observation or estimate and reproduce the desired component through a hear-through secondary-source path. To address these limitations, this paper formulates DP-ANC as a direction-conditioned cancellation--preservation optimization problem. A component-separated objective jointly penalizes residual noise energy and the control response induced by the desired component, with a scalar weighting parameter controlling the cancellation--preservation trade-off. A convolutional network conditioned on the specified desired direction through feature-wise linear modulation (FiLM) is trained using a differentiable secondary-path-aware forward model. At deployment, the network estimates the complete multichannel finite impulse response (FIR) control-filter bank directly from a mixed-reference observation and the specified desired direction in a single forward pass, while retaining the conventional feedforward ANC signal path. Over 3300 evaluation cases, the selected operating point achieves 22.8~dB mean noise reduction with a desired-signal distortion of \textminus 11.4~dB. Validation using measured in-ear-device transfer functions further demonstrates consistent performance under measured acoustic configurations.
\end{abstract}


\IEEEpeerreviewmaketitle
\raggedbottom

\section{Introduction}

Conventional active noise control (ANC) minimizes the total disturbance without distinguishing desired sound from noise \cite{ElliottNelson1993ANC,KuoMorgan1996,Kajikawa2012ANCReview,WenZhang2018OnlineSPM}. Adaptive feedforward implementations, including multichannel FxLMS variants, realize this objective by minimizing the residual error \cite{Burgess1981AdaptiveANC,Elliott1987MultipleErrorLMS,Morgan2013FxLMSHistory,ZhangHongwei,Lian2025FrequencyPoint}. On head-mounted devices, achievable attenuation also depends on feedforward causality, the secondary-path response, and source direction \cite{rafaely2002combined,zhang2014causality}. In personal listening applications, however, it is often necessary to suppress interference arriving from directions other than a specified desired direction while preserving informative sound naturally arriving from that direction \cite{Patel2020DirectionalHearThrough,Zhang2023DirectionalANC,Xiao2023SSANC}. This need motivates direction-preserving ANC (DP-ANC).

Desired-signal preservation methods differ in whether they specify the desired sound by its signal content or by its physical direction of arrival. Signal-component-based methods estimate a target component, such as speech, and reproduce the enhanced signal through the device receiver \cite{DalgaDoclo2011OpenFitting,Serizel2013BinauralANCNR,Doclo2015ALD,VanDenBogaert2009BinauralMWF,Cornelis2010BinauralAnalysis}. The associated processing may introduce algorithmic delay \cite{DalgaDoclo2011OpenFitting,Serizel2013BinauralANCNR} or modify binaural cues \cite{Doclo2015ALD,VanDenBogaert2009BinauralMWF,Cornelis2010BinauralAnalysis}. Direction-based methods instead define all sound arriving from a specified direction as desired. Directional hear-through systems extract this sound by beamforming and reproduce it through secondary sources \cite{Patel2020DirectionalHearThrough,Zhang2023DirectionalANC}. Rather than reconstructing the desired signal, spatially selective ANC (SSANC) imposes a desired-direction response constraint within the ANC optimization, thereby preserving the naturally arriving physical sound \cite{Xiao2023SSANC,XiaoDoclo2024TargetDelay}. A soft-constrained variant trades noise reduction against speech distortion through a scalar parameter while retaining a prescribed desired-response model \cite{Xiao2025SoftSSANC}. Across these formulations, performance depends on the prescribed target response, delay, relative impulse-response model, and regularization \cite{Xiao2023SSANC,XiaoDoclo2024TargetDelay,XiaoDoclo2025Acausal,Xiao2025SoftSSANC}. The need to recompute a matrix solution for each new mixed-reference observation in analytical formulations \cite{Xiao2023SSANC,Xiao2025SoftSSANC} motivates offline control-filter estimation. Efficient multichannel control-filter estimation using structured formulations, such as Kronecker-product decomposition, has also been investigated \cite{LeePark2025KPD,Li2025NKPMFxAP}.

Neural-network-based ANC methods can be broadly grouped by whether the network directly generates the control signal or provides a control filter for a conventional feedforward path. Deep ANC exemplifies direct signal generation and has been studied for broadband and nonlinear control \cite{ZhangWang2021DeepANC,ZhangWang2023DeepMCANC,bai2025wavenet}. Filter-oriented methods include selective fixed-filter ANC, which selects from a bank of pre-trained filters and has been extended with source-direction information \cite{Shi2020SFANC,YangJunSFANC,luo2024SFANC,Wang2025DirectionalSFANC,Shen2022AdaptiveGain}; generative fixed-filter ANC (GFANC), which estimates control filters from reference signals \cite{Luo2023GFANC} and has been extended to incorporate spatial-frequency cues in reverberant environments \cite{Wang2026SpatialFrequencyGFANC}; and hybrid approaches that combine learned filter initialization with subsequent adaptive filtering or joint secondary-path modeling \cite{Luo2026GFANCFxNLMS,Yang2026MetaANC,aboutiman2026hybrid,Christensen}. Most of these methods target noise attenuation alone. None jointly trains a direction-conditioned control-filter estimation network with an explicit cancellation--preservation objective.

Selective fixed-filter ANC chooses from a predesigned control-filter bank but does not generate new filters for unseen conditions or incorporate a preservation objective. GFANC estimates a new control filter from the reference observation but has been applied to noise attenuation without an explicit preservation term. Analytical SSANC enforces a prescribed desired-direction response through a matrix solution recomputed for each new observation, while beamformer hear-through extracts and reproduces the desired component through the secondary-source path. Taken together, none of these approaches produces a control-filter bank that explicitly balances noise cancellation and desired-signal preservation without relying on a predesigned control-filter bank, repeated analytical optimization, or desired-signal reconstruction.

To address this gap, this paper proposes a data-driven DP-ANC method that suppresses the noise component while limiting the control response to sound arriving from a specified desired direction. The framework employs a direction-conditioned neural network to estimate a complete multichannel finite impulse response (FIR) control-filter bank from a short mixed-reference observation. The desired direction is encoded using a periodic representation and injected into each convolutional block through feature-wise linear modulation (FiLM) \cite{Perez2018FiLM}, which has also been applied to secondary-path conditioning in ANC \cite{Yuan2026SmartGlassesANC}. During offline training, the estimated filters are evaluated through a differentiable secondary-path-aware forward model using separated desired and noise components. At deployment, only the mixed-reference observation and the specified desired direction are required, while the estimated filters operate within a conventional feedforward ANC structure. The proposed framework is evaluated using both a simulated circular-array configuration and a separately trained model based on measured transfer functions from the Hearpiece in-ear-device database \cite{Denk2021Hearpiece}.

The main contributions are summarized as follows:
\begin{itemize}
    \item We formulate direction-preserving ANC as a direction-conditioned cancellation--preservation optimization problem, where a scalar weighting parameter continuously balances noise cancellation and desired-signal preservation without imposing a prescribed desired-direction response constraint.

    \item We develop an offline-trained convolutional network that maps a short mixed-reference observation and the specified desired direction directly to the complete multichannel FIR control-filter bank in a single forward pass, while retaining the conventional feedforward ANC signal path.

    \item We characterize the achievable cancellation--preservation region of direction-preserving ANC and show that it is shaped by the spatial separability of the reference array, with consistent trends observed for both learning-based and analytical control-filter designs in simulated and measured acoustic configurations.
\end{itemize}

\section{Signal Model and DP-ANC Formulation}
\label{sec:problem}

\subsection{Multichannel Feedforward ANC Model}

Consider a feedforward ANC system with $K$ reference microphones,
one secondary source, and one error microphone. For an FIR control-filter
length $L_w$, let
\begin{equation}
    \mathbf{x}_k(n)
    =
    [x_k(n),x_k(n-1),\ldots,x_k(n-L_w+1)]^{\mathrm T},
\end{equation}
denote the tapped vector of the $k$-th reference signal, and let
$\mathbf{w}_k\in\mathbb{R}^{L_w}$ denote its control filter. Stacking
these vectors as $\mathbf{x}(n)$ and $\mathbf{w}$ gives the
secondary-source driving signal
\begin{equation}
    y(n)
    =
    \sum_{k=1}^{K}\mathbf{w}_k^{\mathrm T}\mathbf{x}_k(n)
    =
    \mathbf{w}^{\mathrm T}\mathbf{x}(n).
    \label{eq:control_signal_multichannel}
\end{equation}

Let $g(n)$ denote the $L_g$-tap causal secondary-path impulse response and
$d(n)$ the primary disturbance
at the error microphone. Define the secondary-path-filtered references
by $\tilde{x}_k(n) = g(n)\ast x_k(n)$, where $\ast$ denotes linear
convolution, and form their tapped and stacked vector
$\tilde{\mathbf{x}}(n)$ in the same manner as $\mathbf{x}(n)$. The
residual error is then
\begin{equation}
    e(n)
    =
    d(n)+\mathbf{w}^{\mathrm T}\tilde{\mathbf{x}}(n),
\end{equation}
where $\mathbf{w}^{\mathrm T}\tilde{\mathbf{x}}(n)$ denotes the secondary-path-filtered control contribution designed to destructively interfere with $d(n)$.

For a segment of $N$ samples, let
$\mathbf{d},\mathbf{e}\in\mathbb{R}^{N}$ denote the disturbance and
residual vectors, and let
$\tilde{\mathbf{X}}\in\mathbb{R}^{N\times KL_w}$ denote the
secondary-path-filtered reference matrix. Thus,
\begin{equation}
    \mathbf{e}
    =
    \mathbf{d}+\tilde{\mathbf{X}}\mathbf{w}.
    \label{eq:segment_residual}
\end{equation}
Conventional feedforward ANC minimizes the residual energy
\cite{KuoMorgan1996,Elliott2000}
\begin{equation}
    \min_{\mathbf{w}}
    \left\|
    \mathbf{d}
    +
    \tilde{\mathbf{X}}\mathbf{w}
    \right\|_2^2.
    \label{eq:conventional_anc_objective}
\end{equation}
With $\ell_2$ regularization, the corresponding least-squares solution is
\begin{equation}
    \mathbf{w}_{\mathrm{LS}}
    =
    -
    \left(
    \tilde{\mathbf{X}}^{\mathrm T}\tilde{\mathbf{X}}
    +
    \mu\mathbf{I}
    \right)^{-1}
    \tilde{\mathbf{X}}^{\mathrm T}\mathbf{d},
    \label{eq:ls_solution}
\end{equation}
where $\mu\geq0$ is a regularization parameter.

\subsection{Directional Signal Decomposition}

The conventional objective in \eqref{eq:conventional_anc_objective} suppresses the total disturbance without distinguishing desired sound from noise. To formulate the direction-selective objective, we decompose the disturbance and reference signals into directional components. The separated components are used for offline objective evaluation, whereas the deployed control-filter estimation network uses only the mixed-reference observation and specified desired direction. This work considers one desired source and one noise source. Let $\theta_d$ and $\theta_n$ denote their directions, respectively.

For a source signal $u(n)$ arriving from direction $\theta$, the reference signal at the $k$-th microphone and the primary disturbance at the error microphone are modeled as
\begin{equation}
    x_k^{(\theta)}(n)
    =
    q_k(\theta,n)\ast u(n),
    \quad
    d^{(\theta)}(n)
    =
    p(\theta,n)\ast u(n),
    \label{eq:directional_paths}
\end{equation}
where $q_k(\theta,n)$ and $p(\theta,n)$ are the corresponding acoustic paths.

For the two-source case,
\begin{equation}
    \mathbf{d}
    =
    \mathbf{d}_n+\mathbf{d}_d,
    \quad
    \tilde{\mathbf{X}}
    =
    \tilde{\mathbf{X}}_n+\tilde{\mathbf{X}}_d,
    \label{eq:linear_decomposition}
\end{equation}
where the subscripts $n$ and $d$ denote the noise and desired components. Hereafter, $\mathbf{d}_n$ together with $\tilde{\mathbf{X}}_n$ is referred to as the noise component, and $\mathbf{d}_d$ together with $\tilde{\mathbf{X}}_d$ as the desired component. By linearity,
\begin{equation}
    \mathbf{e}
    =
    \underbrace{
    \left(
    \mathbf{d}_n
    +
    \tilde{\mathbf{X}}_n\mathbf{w}
    \right)
    }_{\text{residual noise component}}
    +
    \underbrace{
    \left(
    \mathbf{d}_d
    +
    \tilde{\mathbf{X}}_d\mathbf{w}
    \right)
    }_{\text{desired component after control}}.
    \label{eq:residual_decomposition}
\end{equation}
Noise attenuation requires
\begin{equation}
    \mathbf{d}_n
    +
    \tilde{\mathbf{X}}_n\mathbf{w}
    \approx
    \mathbf{0},
    \label{eq:noise_reduction_requirement}
\end{equation}
whereas desired-signal preservation requires
\begin{equation}
    \mathbf{d}_d
    +
    \tilde{\mathbf{X}}_d\mathbf{w}
    \approx
    \mathbf{d}_d,
    \quad
    \text{or equivalently}
    \quad
    \tilde{\mathbf{X}}_d\mathbf{w}
    \approx
    \mathbf{0}.
    \label{eq:desired_preservation_requirement}
\end{equation}
Thus, the same control-filter bank must cancel the noise component while generating little secondary response to the desired component.

\begin{figure*}[!t]
    \centering
    \includegraphics[width=0.95\textwidth]{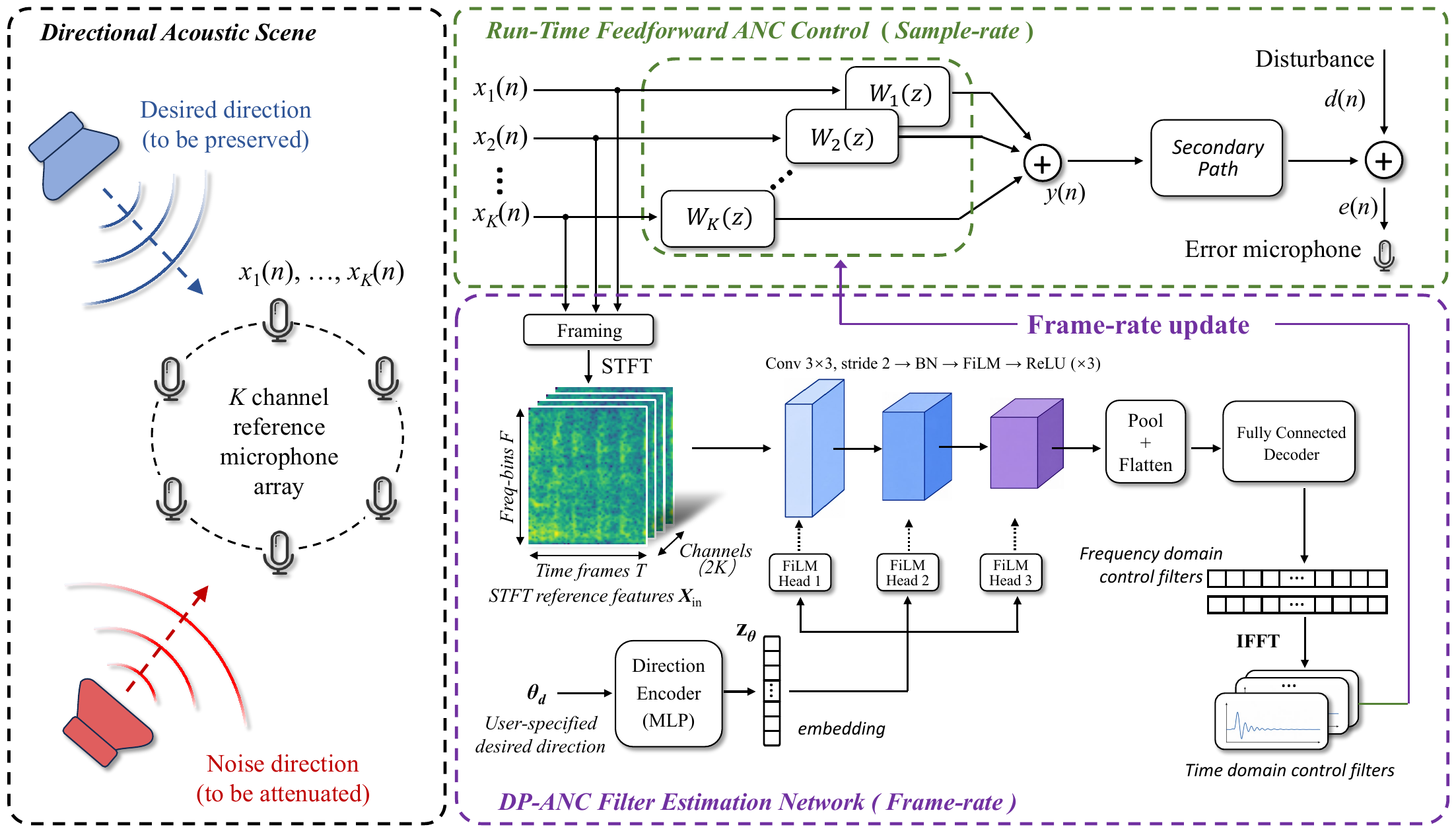}
    \caption{Overview of the proposed DP-ANC framework. During offline training, multichannel reference features and the specified desired direction are mapped to a feedforward control-filter bank and evaluated through a differentiable secondary-path-aware forward model. At runtime, the estimated filters are used in the retained feedforward ANC structure to attenuate the noise component while limiting the control response to the desired component.}
    \label{fig:overall_framework}
\end{figure*}

\subsection{Cancellation--Preservation Objective}

The two requirements are incorporated into a component-separated objective. Imposing the exact broadband equality $\tilde{\mathbf{X}}_d\mathbf{w} = \mathbf{0}$ can substantially reduce the filter degrees of freedom available for noise cancellation. We therefore penalize the energy of the desired-induced control response rather than constraining it to zero, yielding a continuous trade-off without prescribing a target response or delay.

To reduce dependence on source level, define
\begin{equation}
    \alpha_n
    =
    \|\mathbf{d}_n\|_2^2+\epsilon,
    \qquad
    \alpha_d
    =
    \|\mathbf{d}_d\|_2^2+\epsilon,
    \label{eq:normalization_energies}
\end{equation}
and
\begin{equation}
    \bar{\mathbf{d}}_n
    =
    \frac{\mathbf{d}_n}{\sqrt{\alpha_n}},
    \qquad
    \bar{\mathbf{X}}_n
    =
    \frac{\tilde{\mathbf{X}}_n}{\sqrt{\alpha_n}},
    \qquad
    \bar{\mathbf{X}}_d
    =
    \frac{\tilde{\mathbf{X}}_d}{\sqrt{\alpha_d}},
    \label{eq:normalized_directional_terms}
\end{equation}
where $\epsilon$ is a small positive constant. All terms are evaluated after restriction to the processing band.

The normalized noise-reduction (NR) term is
\begin{equation}
    \mathcal{L}_{\mathrm{NR}}(\mathbf{w})
    =
    \left\|
    \bar{\mathbf{d}}_n
    +
    \bar{\mathbf{X}}_n\mathbf{w}
    \right\|_2^2,
    \label{eq:nr_loss}
\end{equation}
and the desired-signal preservation term is
\begin{equation}
    \mathcal{L}_{\mathrm{PR}}(\mathbf{w})
    =
    \left\|
    \bar{\mathbf{X}}_d\mathbf{w}
    \right\|_2^2.
    \label{eq:pr_loss}
\end{equation}
The training objective is
\begin{equation}
    \mathcal{L}_{\mathrm{DP}}(\mathbf{w})
    =
    \mathcal{L}_{\mathrm{NR}}(\mathbf{w})
    +
    \lambda
    \mathcal{L}_{\mathrm{PR}}(\mathbf{w}),
    \label{eq:dp_loss}
\end{equation}
where $\lambda\geq0$ controls the trade-off. Larger values place more weight on limiting the desired-induced control response, whereas smaller values permit more aggressive cancellation. Related weighted noise-reduction and distortion trade-offs appear in speech-distortion-weighted multichannel Wiener filtering \cite{Spriet2004SDWMWF,DocloMoonen2005OutputSNR,Doclo2007SDWMWF}. Soft-constrained SSANC likewise uses a scalar parameter to trade noise reduction against speech distortion while retaining a prescribed desired-response model and obtaining the control filters through an analytical matrix solution \cite{Xiao2025SoftSSANC}. In contrast, \eqref{eq:dp_loss} forms its cancellation and preservation terms from separated noise and desired components at the ANC error microphone, with the control responses evaluated through the secondary path. The preservation term penalizes the desired-induced control response rather than deviation from a prescribed desired-direction response.

Analytical SSANC minimizes total output power while enforcing a prescribed desired-direction response through either a hard constraint or a soft penalty. Its performance depends on the target response, delay, path model, covariance estimate, and regularization \cite{Xiao2023SSANC,XiaoDoclo2024TargetDelay,XiaoDoclo2025Acausal,Xiao2025SoftSSANC}. The formulations therefore define their operating trade-offs differently, motivating the frontier comparison in Section~\ref{subsec:lambda_ablation}.

A regularized closed-form counterpart of \eqref{eq:dp_loss} clarifies how the preservation term shapes the control-filter solution. Consider
\begin{equation}
    J_{\mathrm{DP}}(\mathbf{w})
    =
    \mathcal{L}_{\mathrm{NR}}(\mathbf{w})
    +
    \lambda
    \mathcal{L}_{\mathrm{PR}}(\mathbf{w})
    +
    \mu\|\mathbf{w}\|_2^2.
    \label{eq:weighted_objective}
\end{equation}
Its normal equation is
\begin{equation}
    \left(
    \bar{\mathbf{X}}_n^{\mathrm T}\bar{\mathbf{X}}_n
    +
    \lambda
    \bar{\mathbf{X}}_d^{\mathrm T}\bar{\mathbf{X}}_d
    +
    \mu\mathbf{I}
    \right)\mathbf{w}
    =
    -\bar{\mathbf{X}}_n^{\mathrm T}\bar{\mathbf{d}}_n,
    \label{eq:weighted_normal_equation}
\end{equation}
which gives, when the matrix is nonsingular,
\begin{equation}
    \mathbf{w}_{\lambda}^{\star}
    =
    -
    \left(
    \bar{\mathbf{X}}_n^{\mathrm T}\bar{\mathbf{X}}_n
    +
    \lambda
    \bar{\mathbf{X}}_d^{\mathrm T}\bar{\mathbf{X}}_d
    +
    \mu\mathbf{I}
    \right)^{-1}
    \bar{\mathbf{X}}_n^{\mathrm T}\bar{\mathbf{d}}_n.
    \label{eq:weighted_solution}
\end{equation}
Because it depends on the desired direction, the second Gram matrix acts as a quadratic regularizer that penalizes filters producing a strong control response to the desired component. Direct evaluation of \eqref{eq:weighted_solution} requires the separated components and a new solution of the normal equations for each input observation. The learning-based realization in the next subsection instead uses \eqref{eq:dp_loss} with $\mu = 0$ for direct training. Section~\ref{subsec:network_training_configuration} gives the optimization settings.

\section{Proposed DP-ANC Framework}
\label{sec:framework}

\subsection{Framework Overview}

Fig.~\ref{fig:overall_framework} summarizes the learning-based realization of the component-separated objective. Let $\mathbf{X}\in\mathbb{R}^{K\times N_{\mathrm{obs}}}$ denote a mixed-reference observation comprising $K$ time-domain channels. Its complex short-time Fourier transform (STFT) forms the network input $\mathbf{X}_{\mathrm{in}}$ defined in Section~\ref{subsec:filter_estimator}. Given $\mathbf{X}$ and the specified desired direction, the estimation network produces the complete feedforward control-filter bank:
\begin{equation}
    \hat{\mathbf{W}}
    =
    F_\Theta(\mathbf{X},\theta_d)
    \in \mathbb{R}^{K\times L_w}.
    \label{eq:framework_mapping}
\end{equation}
The network does not receive the noise direction $\theta_n$ explicitly. The multichannel magnitude and phase relationships in $\mathbf{X}_{\mathrm{in}}$ encode the incident spatial structure, while $\theta_d$ identifies the component whose control response is to be limited. The desired direction is supplied as an external input because direction-of-arrival estimation is not part of the present network.

The separated components required by \eqref{eq:dp_loss} are available in the generated training scenes and are used only to evaluate the offline objective. The network parameters are obtained as
\begin{equation}
    \Theta^{\star}
    =
    \arg\min_{\Theta}
    \mathbb{E}_{\mathcal{T}}
    \left[
    \mathcal{L}_{\mathrm{DP}}
    \left(
    F_{\Theta}(\mathbf{X},\theta_d)
    \right)
    \right],
    \label{eq:amortized_training}
\end{equation}
where $\mathcal{T}$ denotes the distribution of training scenes over directions, paths, source levels, and waveform realizations. Minimizing this expectation learns a single mapping shared across these training scenes, replacing the matrix solve that would otherwise be repeated for each deployment observation.

The convolutional encoder is followed by adaptive pooling, allowing the same model to accept different observation durations. Once the control-filter bank is available, the control signal is
\begin{equation}
    y(n)
    =
    \sum_{k=1}^K
    \hat{\mathbf{w}}_k^{\mathrm T}\mathbf{x}_k(n),
    \label{eq:runtime_filtering_section3}
\end{equation}
and is reproduced through the secondary source. Deployment therefore requires neither separated desired and noise components nor an explicit noise-direction label.

\subsection{Direction-Conditioned Control-Filter Estimation Network}
\label{subsec:filter_estimator}

The real and imaginary parts of the $K$ reference-microphone STFTs are stacked to form
\begin{equation}
    \mathbf{X}_{\mathrm{in}}
    \in
    \mathbb{R}^{2K\times n_f\times T},
    \label{eq:stft_input}
\end{equation}
where $n_f$ and $T$ denote the number of frequency bins and frames. Retaining the complex multichannel representation preserves the inter-microphone phase relationships required for directional discrimination \cite{Torres2012CircularArray}.

The desired direction is represented by the periodic code
\begin{equation}
    \mathbf{c}(\theta_d)
    =
    [\cos\theta_d,\sin\theta_d]^{\mathrm T},
    \label{eq:direction_code}
\end{equation}
and embedded by a two-layer direction encoder,
\begin{equation}
    \mathbf{z}_{\theta}
    =
    E_d\!\left(\mathbf{c}(\theta_d)\right)
    \in\mathbb{R}^{32},
    \label{eq:direction_encoder}
\end{equation}
where both linear layers have 32 outputs and use sigmoid linear unit (SiLU) activations.

Convolution (Conv), batch normalization (BN), layer normalization (LN), and rectified linear unit (ReLU) are abbreviated below and in Table~\ref{tab:netcfg}.
Let $\mathbf{u}_{\ell}$ denote the output of convolution and batch normalization in acoustic encoder block $\ell$. A block-specific FiLM head \cite{Perez2018FiLM} maps the direction embedding to channel-wise modulation parameters:
\begin{equation}
    [\boldsymbol{\gamma}_{\ell}^{\mathrm T},
    \boldsymbol{\beta}_{\ell}^{\mathrm T}]^{\mathrm T}
    =
    A_{\ell}(\mathbf{z}_{\theta}).
    \label{eq:film_parameters}
\end{equation}
For a block with $C_{\ell}$ output channels, both
$\boldsymbol{\gamma}_{\ell}$ and $\boldsymbol{\beta}_{\ell}$ belong to
$\mathbb{R}^{C_{\ell}}$. They are broadcast over frequency and time, and
the block output is
\begin{equation}
    \mathbf{h}_{\ell}
    =
    \operatorname{ReLU}\!\left[
    (\mathbf{1}+\boldsymbol{\gamma}_{\ell})
    \odot\mathbf{u}_{\ell}
    +\boldsymbol{\beta}_{\ell}
    \right].
    \label{eq:film_modulation}
\end{equation}
The modulation in \eqref{eq:film_modulation} is applied after BN and before
ReLU in each of the three convolutional blocks. All weights and biases in
$A_{\ell}$ are initialized to zero, so that
$\boldsymbol{\gamma}_{\ell} = \boldsymbol{\beta}_{\ell} = \mathbf{0}$ at
initialization and the initial modulation is an identity mapping.

After the third FiLM block, adaptive pooling and flattening produce the
acoustic representation $\mathbf{z}_x\in\mathbb{R}^{2048}$. The decoder maps
this representation to the normalized real and imaginary coefficients of all
$K$ control filters:
\begin{equation}
    \hat{\mathbf{H}}_{\mathrm{norm}}
    =
    D_w(\mathbf{z}_x)
    \in\mathbb{R}^{2Kn_f}.
    \label{eq:conditioned_filter_decoder}
\end{equation}
The architecture is summarized in Table~\ref{tab:netcfg}. Adaptive pooling to $8\times4$ retains a
compact frequency--time description, while the output layer provides separate
coefficients for each frequency bin and reference channel. The parameter count
for each acoustic encoder block includes its FiLM head.

\begin{table*}[!t]
\centering
\caption{Layer-wise architecture of the FiLM-conditioned control-filter estimation network. The main simulated configuration uses $K = 6$, $n_f = 129$, and $L_w = 256$, with $T$ denoting the number of STFT frames.}
\label{tab:netcfg}
\setlength{\tabcolsep}{3.5pt}
\renewcommand{\arraystretch}{1.08}
\begin{tabular}{@{}llllr@{}}
\toprule
Stage & Layer & Configuration & Output shape & Parameters \\
\midrule
Input
& STFT (real/imaginary)
& $256$-point transform, hop $64$
& $12\times129\times T$
& -- \\
\midrule
\multirow{2}{*}{Direction encoder ($E_d$)}
& Linear--SiLU
& $2\to32$
& $32$
& 96 \\
& Linear--SiLU
& $32\to32$
& $32$
& 1,056 \\
\midrule
\multirow{5}{*}{Acoustic encoder}
& Conv--BN--FiLM--ReLU
& $3{\times}3$, stride 2, $C_1 = 16$
& $16\times65\times\lceil T/2\rceil$
& 2,832 \\
& Conv--BN--FiLM--ReLU
& $3{\times}3$, stride 2, $C_2 = 32$
& $32\times33\times\lceil T/4\rceil$
& 6,816 \\
& Conv--BN--FiLM--ReLU
& $3{\times}3$, stride 2, $C_3 = 64$
& $64\times17\times\lceil T/8\rceil$
& 22,848 \\
& Adaptive average pooling
& $8\times4$
& $64\times8\times4$
& -- \\
& Flatten
& --
& $2048$
& -- \\
\midrule
\multirow{2}{*}{Decoder ($D_w$)}
& Linear--LN--ReLU--Dropout
& $2048\to256$, $p = 0.15$
& $256$
& 525,056 \\
& Linear
& $256\to 2n_fK$
& $1548$
& 397,836 \\
\midrule
Reconstruction
& de-normalization + inverse transform
& Hermitian completion, $256$-point IFFT
& $6\times256$
& -- \\
\midrule
Total
& --
& --
& --
& 956,540 \\
\bottomrule
\end{tabular}
\end{table*}

The estimated coefficients are de-normalized using training-set statistics and reshaped into $K$ one-sided complex spectra. Each spectrum is Hermitian-completed to 256 bins and transformed into a real causal FIR filter:
\begin{equation}
    \hat{\mathbf{w}}_k
    =
    \mathrm{IFFT}_{256}\{\hat{\mathbf{H}}_k\}
    \in\mathbb{R}^{L_w},
    \quad k=1,\ldots,K.
    \label{eq:ifft_filter}
\end{equation}
Here, $\hat{\mathbf{H}}_k\in\mathbb{C}^{256}$ denotes the Hermitian-completed spectrum and $L_w = 256$. Thus, the transform directly yields the complete FIR filter without truncation. Frequency-domain estimation is used only as a compact filter parameterization, while runtime control remains time-domain feedforward convolution.

\subsection{Differentiable Training with Independent Signal Realizations}
\label{subsec:independent_realizations}

Each training sample uses a mixed-reference observation for filter estimation (realization A) and independent signals for loss evaluation (realization B). The two realizations share the same source directions and acoustic paths but use independent source waveforms. This separation encourages the estimated filters to remain effective beyond the particular waveforms used for filter estimation.

During training, the differentiable secondary-path-aware forward model applies the vectorized estimated control-filter bank $\hat{\mathbf{w}}$ separately to the secondary-path-filtered noise and desired-component matrices $\tilde{\mathbf{X}}_n$ and $\tilde{\mathbf{X}}_d$ formed from the independent signals used for loss evaluation. After band limitation and the normalization in \eqref{eq:normalized_directional_terms}, the resulting control responses enter the cancellation and desired-signal preservation terms in \eqref{eq:nr_loss} and \eqref{eq:pr_loss}. Their gradients are propagated through the control filters and secondary path.

\section{Experimental Setup}
\label{sec:experimental_setup}

\subsection{Main Simulated Acoustic Configuration}
\label{subsec:main_simulated_configuration}

\begin{figure}[!t]
\centering
\includegraphics[width=0.80\columnwidth]{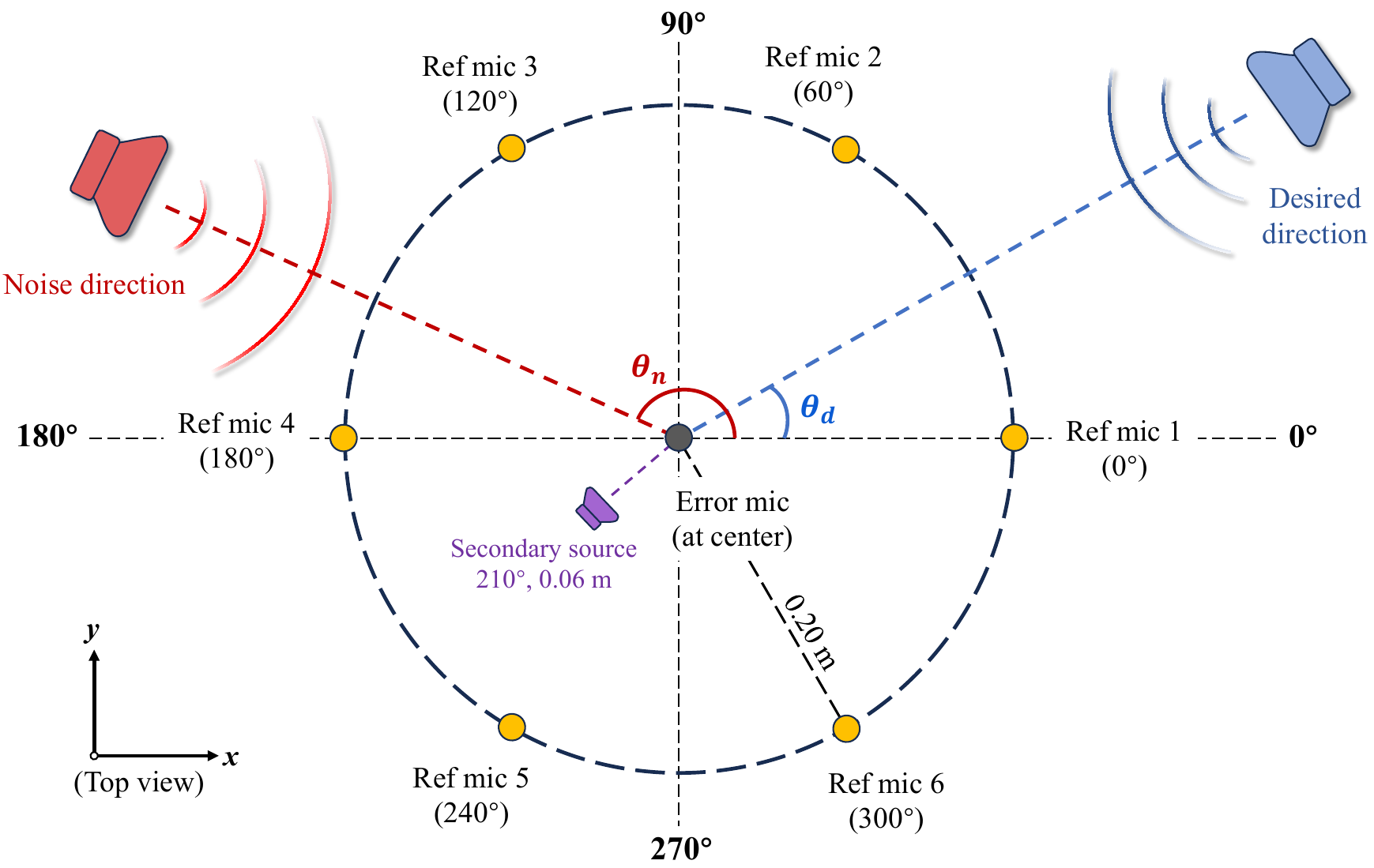}
\caption{Main simulated acoustic configuration. Six reference microphones surround the center error microphone, and one secondary source is located at $210^{\circ}$ and a radius of $0.06$~m.}
\label{fig:simulated_configuration}
\end{figure}

The main experiments use the multichannel feedforward arrangement in Fig.~\ref{fig:simulated_configuration}. Six reference microphones form a uniform circle around the central error microphone, with one secondary source located nearby. Horizontal plane-wave propagation is modeled by direction-dependent delays across the circular array \cite{Tiana2010CircularArray,Torres2012CircularArray}. These delays are implemented using fractional-delay FIR filters \cite{Laakso1996FractionalDelay}. The sampling rate is 8~kHz, the processing band is 20--2500~Hz, and each filter estimate uses a 0.5~s reference observation.

The control filters contain 256 taps per reference channel, and the secondary path is represented by a 16-tap causal response that is treated as known and held fixed in training and evaluation. In the main evaluation, independent band-limited sensor noise is added to all reference and error-microphone channels. Its level is set independently in each channel to yield a 30~dB sensor SNR relative to the local observed mixture. The evaluation uses a pre-control SNR of \textminus{}10~dB at the error microphone and angular separations $\Delta\theta\geq15^\circ$. Table~\ref{tab:main_configuration} summarizes the acoustic and optimization settings.

To provide a geometry-based reference for the main separation criterion and the near-direction experiment, we characterize the spatial separability of the circular reference array. Its far-field response at microphone $k$ is
\begin{equation}
    Q_k(\theta,f)
    =
    \exp\left[-j\kappa r\cos(\theta-\phi_k)\right],
    \qquad
    \kappa=\frac{2\pi f}{c},
    \label{eq:uca_directional_response}
\end{equation}
where $r$ is the array radius, $\phi_k = 2\pi(k-1)/K$, and $c$ is the speed of sound \cite{Tiana2010CircularArray,Torres2012CircularArray}. Defining $\mathbf{q}(\theta,f) = [Q_1(\theta,f),\ldots,Q_K(\theta,f)]^{\mathrm T}$, the normalized manifold coherence is
\begin{equation}
\begin{split}
    \eta(f,\Delta\theta)
    &=
    \frac{\left|\mathbf{q}^{\mathrm H}(\theta_d,f)\mathbf{q}(\theta_n,f)\right|}
    {\|\mathbf{q}(\theta_d,f)\|_2\|\mathbf{q}(\theta_n,f)\|_2}\\
    &\approx
    \left|J_0\left(2\kappa r\sin\frac{\Delta\theta}{2}\right)\right|,
\end{split}
\label{eq:uca_manifold_coherence}
\end{equation}
where $\theta_n = \theta_d+\Delta\theta$ and the second line is the continuous-circular-aperture approximation \cite{Tiana2010CircularArray}. For unit-norm responses, $1-\eta^2$ is the fraction of the noise-response energy orthogonal to the desired-response direction. Large coherence therefore limits the simultaneous achievement of high NR and low desired-signal distortion.

Under the conventional half-power coherence criterion, $\eta^2 = 1/2$ \cite{Tiana2010CircularArray}, the corresponding angular separation is
\begin{equation}
    \Delta\theta_{\mathrm{3dB}}(f)
    =
    2\sin^{-1}\left(\frac{1.126c}{4\pi fr}\right),
    \label{eq:uca_separability_angle}
\end{equation}
where 1.126 is the first positive solution of $J_0(x) = 1/\sqrt{2}$. For $r = 0.20$~m, the approximation gives frequency-dependent resolution scales of about $36^\circ$, $18^\circ$, $12^\circ$, and $9^\circ$ at 500, 1000, 1500, and 2000~Hz. At high frequencies, finite six-sensor sampling and discrete-array spatial aliasing also influence the response \cite{Torres2012CircularArray}. These scales provide an interpretive reference for the angular-resolution experiment, but the $\Delta\theta \geq 15^\circ$ main-evaluation criterion is not a frequency-independent physical resolution boundary.

\begin{table*}[!t]
\centering
\caption{Main acoustic and optimization settings.}
\label{tab:main_configuration}
\setlength{\tabcolsep}{5.5pt}
\renewcommand{\arraystretch}{1.08}
\begin{tabular}{@{}llll@{}}
\toprule
\multicolumn{2}{c}{\textbf{Acoustic configuration}}
&
\multicolumn{2}{c}{\textbf{Optimization configuration}}
\\
\cmidrule(r){1-2}
\cmidrule(l){3-4}
Parameter & Value & Parameter & Value \\
\midrule
Array geometry
& $K = 6$, $r = 0.20$~m, error mic at center
& Preservation weights
& $\{0,0.03,0.1,0.3,1\}$
\\
Secondary source
& $210^{\circ}$, $r = 0.06$~m
& Optimizer
& AdamW
\\
Sampling rate / band
& $8$~kHz / $20$--$2500$~Hz
& Initial learning rate / schedule
& $2\times10^{-4}$ / cosine decay
\\
Reference-signal duration
& $0.5$~s
& Weight decay
& $5\times10^{-4}$
\\
Filter lengths
& $L_w = 256$, $L_g = 16$
& Batch size
& $128$
\\
Training / evaluation SNR
& $\mathcal{U}[-15,5]$~dB / \textminus{}10~dB
& Gradient clipping
& $0.5$
\\
Sensor SNR (per channel)
& $30$~dB
& Validation-set size
& $1024$ scenes
\\
Desired/noise training azimuths
& Independent $\mathcal{U}[0^\circ,360^\circ)$ draws
& Representative preservation weight
& $\lambda = 0.1$
\\
\bottomrule
\end{tabular}
\vspace{2pt}

\parbox{\textwidth}{\footnotesize\textit{Note:} Sensor SNR is computed independently in each channel from the standard-deviation ratio of the local observed mixture to the added sensor noise.}
\end{table*}

\subsection{Data Generation and Directional Sampling}
\label{subsec:scene_generation}

Each training scene contains one desired source and one noise source. Their azimuths $\theta_d$ and $\theta_n$ are drawn independently from continuous uniform distributions over $[0^\circ,360^\circ)$ and remain fixed within each 0.5~s realization.

The desired waveform is a band-limited Gaussian sequence that provides broadband excitation. The held-out experiments in Section~\ref{subsec:evaluation_protocol} test generalization to recorded desired signals. The noise class is sampled with equal probability between a band-limited Gaussian sequence and an UrbanSound8K recording \cite{Salamon2014UrbanSound}. Non-overlapping recording splits are used for training and validation. All recordings are resampled, normalized, and restricted to the processing band before spatial rendering.

During training, the pre-control SNR at the error microphone is sampled independently for each scene from $\mathcal{U}[-15,5]$~dB. The independent-signal protocol in Section~\ref{subsec:independent_realizations} is used throughout training, validation, and evaluation, with separate waveform realizations for filter estimation and loss or metric computation.

\subsection{Network Training}
\label{subsec:network_training_configuration}

We train separate instances of the 956,540-parameter estimation network in Table~\ref{tab:netcfg} for $\lambda\in\{0,0.03,0.1,0.3,1\}$. For each model, we select the training epoch that yields the lowest $\mathcal{L}_{\mathrm{DP}}$ on a fixed validation set of 1024 synthetically generated scenes. All models use the same validation set, generated using the training-scene protocol with separate recordings. After selecting the best-performing epoch for each model, NR and $-D_{\mathrm{des}}$ are independently min--max normalized across the five candidates. The candidate with the smallest Euclidean distance to the ideal normalized point $(1,1)$ defines the representative operating point, yielding $\lambda = 0.1$.

AdamW \cite{LoshchilovHutter2019AdamW} optimizes each model using the hyperparameters summarized in Table~\ref{tab:main_configuration}.

\subsection{Compared Methods}
\label{subsec:compared_methods}

\subsubsection{ANC Off}
The uncontrolled error-microphone signal defines the NR reference at 0~dB.

\subsubsection{Conventional Wiener ANC}
The unconstrained Wiener solution minimizes the total error-microphone energy without distinguishing desired and noise components. It represents the maximum-cancellation endpoint of the conventional objective.

\subsubsection{Beamformer Hear-Through}
We adapt the hear-through signal flow in \cite{Patel2020DirectionalHearThrough} to the present array. A 384-tap least-squares beamformer with a 4.0~ms target delay extracts and reproduces the desired-direction signal through the secondary source after ANC attenuation.

\subsubsection{Analytical SSANC}
For each evaluation case, we recompute the analytical control filters from its 0.5~s mixed-reference observation following the optimal spatially constrained SSANC formulation in \cite{Xiao2023SSANC,XiaoDoclo2024TargetDelay}. The control filters are obtained using the mixed-reference observation and desired-direction path information, without separated source waveforms. The desired response is defined at the error microphone with zero additional delay. Let $\widetilde{\mathbf{x}}(n)$ be the secondary-path-filtered tapped-reference vector and $\mathbf{R}_{\tilde{x}\tilde{x}} = E\{\widetilde{\mathbf{x}}(n)\widetilde{\mathbf{x}}^{\mathrm T}(n)\}$ its covariance. With the block secondary-path matrix $\mathbf{G}$ and desired-direction relative-response matrix $\mathbf{H}$, define
\begin{equation}
    \boldsymbol{\Phi}_{\tilde{x}\tilde{x}}
    =
    \mathbf{R}_{\tilde{x}\tilde{x}}+\beta_{\mathrm{S}}\mathbf{I},
    \qquad
    \mathbf{Q}
    =
    \mathbf{H}^{\mathrm T}\mathbf{G}
    \boldsymbol{\Phi}_{\tilde{x}\tilde{x}}^{-1}
    \mathbf{G}^{\mathrm T}\mathbf{H}.
    \label{eq:ssanc_matrices}
\end{equation}
The covariance and constraint loadings are normalized by the largest eigenvalues of the matrices before the corresponding diagonal loading is added:
\begin{equation}
    \beta_{\mathrm{S}}
    =
    \frac{\lambda_{\max}(\mathbf{R}_{\tilde{x}\tilde{x}})}{r_{\beta}},
    \qquad
    \rho_{\mathrm{S}}
    =
    \frac{\lambda_{\max}(\mathbf{Q})}{r_f},
    \label{eq:ssanc_loadings}
\end{equation}
where $\mathbf{Q}+\rho_{\mathrm{S}}\mathbf{I}$ is the regularized constraint matrix, and $r_{\beta}$ and $r_f$ are dimensionless loading ratios. We fix $r_{\beta} = 10^4$ and sweep $r_f$ over 13 values spanning $2$ to $50{,}000$ to form the SSANC cancellation--preservation frontier. The same 0.5~s observation, 256-tap control filter per channel, and processing band are used at every point. Table~\ref{tab:multicontent_results} and Fig.~\ref{fig:overall_performance} use the representative setting $r_{\beta} = r_f = 10^4$, and method-level trade-off comparison uses the complete frontier.

\subsubsection{DP-ANC (Proposed)}
The proposed estimation network produces the complete control-filter bank from the mixed-reference observation and specified desired direction. The network does not require a new matrix solution for each observation. Its estimated filters operate in the conventional feedforward control path.

\emph{Computational setup.} Operation counts in Section~\ref{subsec:efficiency_results} refer to one control-filter bank obtained from the common 0.5~s observation. For DP-ANC, we count the multiply--accumulate operations (MACs) of the convolutional and fully connected layers. For analytical SSANC, the floating-point operation counts of the eigendecompositions and matrix inversions are converted to MACs using one MAC equal to two floating-point operations.

\subsection{Evaluation Metrics}
\label{subsec:evaluation_metrics}

\begin{figure*}[!t]
\centering
\includegraphics[width=0.95\textwidth]{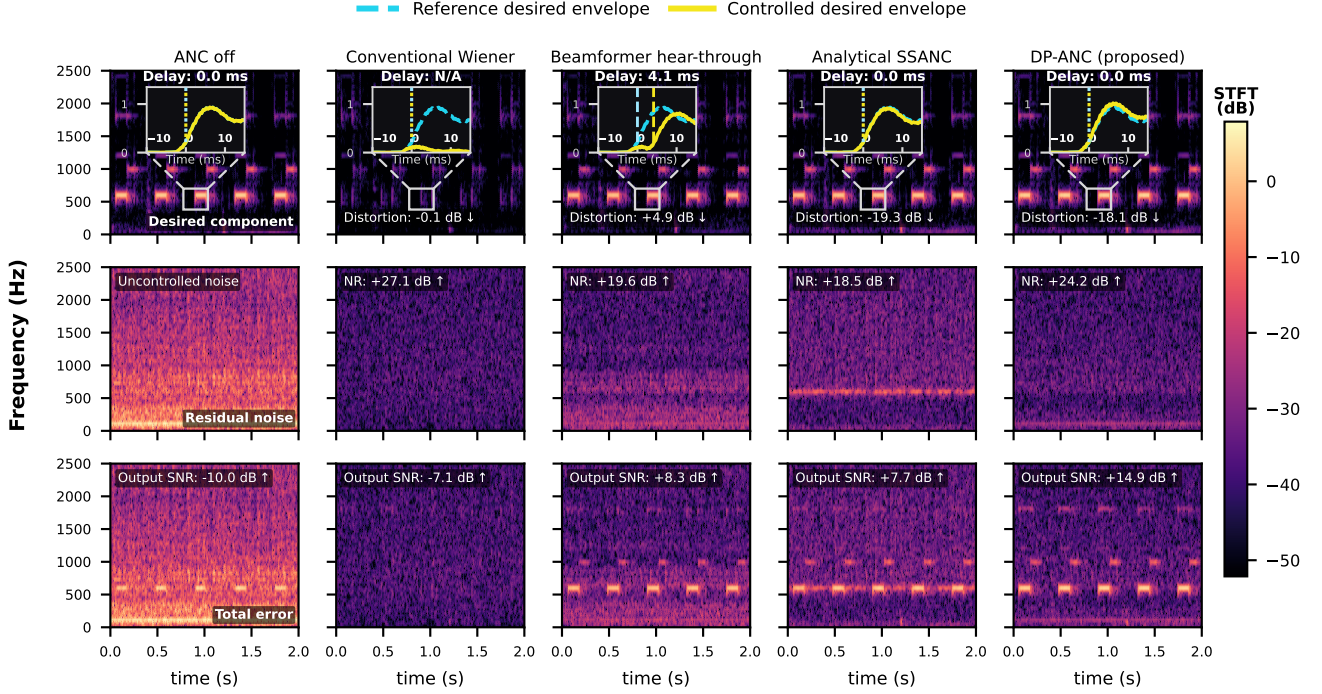}
\caption{Illustrative fixed-record comparison using a tonal alarm at $0^\circ$ and air-conditioner noise at $90^\circ$, with a pre-control SNR of \textminus{}10~dB. Control filters are estimated from the first 0.5~s, and the complete 2.0~s record provides the displayed signals.}
\label{fig:overall_performance}
\end{figure*}

Noise reduction is defined as
\begin{equation}
\mathrm{NR}
=
10\log_{10}
\frac{\sum_n d_n^2(n)}{\sum_n e_n^2(n)},
\label{eq:metric_nr}
\end{equation}
where $d_n(n)$ and $e_n(n)$ are the noise components before and after control. Desired-signal distortion is
\begin{equation}
D_{\mathrm{des}}
=
10\log_{10}
\frac{\sum_n [e_d(n)-d_d(n)]^2}{\sum_n d_d^2(n)},
\label{eq:metric_distortion}
\end{equation}
where $d_d(n)$ and $e_d(n)$ are the desired components before and after control. Consistent with the preservation requirement in \eqref{eq:desired_preservation_requirement}, $D_{\mathrm{des}}$ measures the energy of the desired-induced control response normalized by the naturally arriving desired-component energy and should not be interpreted as a perceptual distortion score. More negative values indicate a weaker response, whereas 0~dB indicates equal energies without distinguishing cancellation, reinforcement, or phase alteration. The output SNR is
\begin{equation}
\mathrm{SNR}_{\mathrm{out}}
=
10\log_{10}
\frac{\sum_n e_d^2(n)}{\sum_n e_n^2(n)}.
\label{eq:metric_output_snr}
\end{equation}
It measures the energy balance between the controlled desired component and residual noise, but does not by itself measure waveform fidelity. We therefore interpret $D_{\mathrm{des}}$ and $\mathrm{SNR}_{\mathrm{out}}$ jointly with the component-resolved time--frequency and envelope comparison in Fig.~\ref{fig:overall_performance}. In radial plots of preservation, $-D_{\mathrm{des}}$ is used only to make a larger radius indicate a weaker desired-induced control response.
For hear-through, delay is estimated by cross-correlation, and desired-component correlation is computed after temporal alignment.

\subsection{Evaluation Protocol}
\label{subsec:evaluation_protocol}

The main evaluation contains 3300 cases in 11 balanced content groups. Six groups pair Gaussian desired signals with recorded noise from six held-out Freesound Dataset 50K (FSD50K) categories \cite{Fonseca2022FSD50K}. The remaining five groups use Gaussian signals, FSD50K alarm, car-horn, or siren recordings, or LibriSpeech test-clean speech \cite{Panayotov2015LibriSpeech} as the desired content, each paired with independently selected recorded noise. The recorded evaluation clips are excluded from network training.

All cases use the main geometry, a \textminus{}10~dB pre-control SNR calibrated at the error microphone over 20--2500~Hz, and angular separations $\Delta\theta\geq15^\circ$. The estimation network uses a 0.5~s mixed-reference observation to estimate the filter, while all reported metrics use an independent 0.5~s signal realization with the same source directions and acoustic paths. The energy sums in \eqref{eq:metric_nr}--\eqref{eq:metric_output_snr} are evaluated over the complete evaluation record after a common 20--2500~Hz analysis filter.

Independent sensor noise is added to the filter-estimation observation and the evaluation signals. DP-ANC and all baselines share the same noisy observation and independent evaluation signals. NR and $\mathrm{SNR}_{\mathrm{out}}$ include source noise and sensor noise in the non-desired component, whereas $D_{\mathrm{des}}$ is computed from the separately rendered desired component.

Table entries report the mean and standard deviation of the per-case metrics in dB.

\begin{figure*}[!t]
\centering
\includegraphics[width=0.92\textwidth]{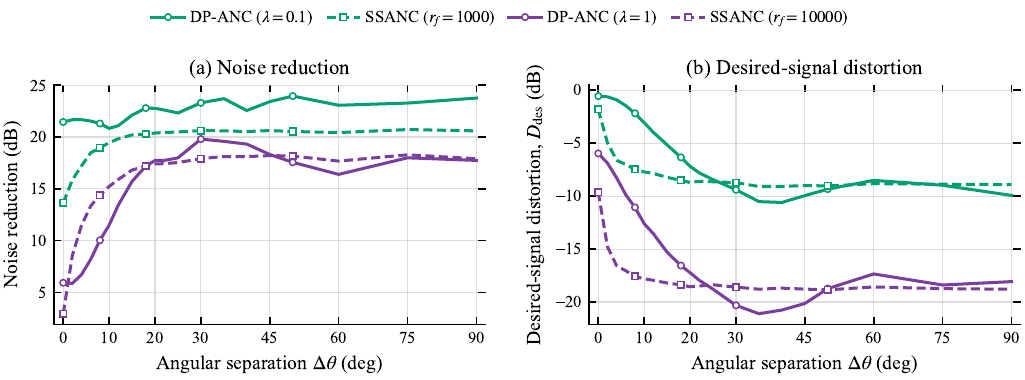}
\caption{Angular-resolution comparison of DP-ANC and analytical SSANC at two cancellation--preservation settings as desired--noise separation varies from $0^\circ$ to $90^\circ$. Each point averages 64 independent cases.}
\label{fig:angular_resolution}
\end{figure*}

\begin{figure*}[t]
\centering
\includegraphics[width=0.98\textwidth]{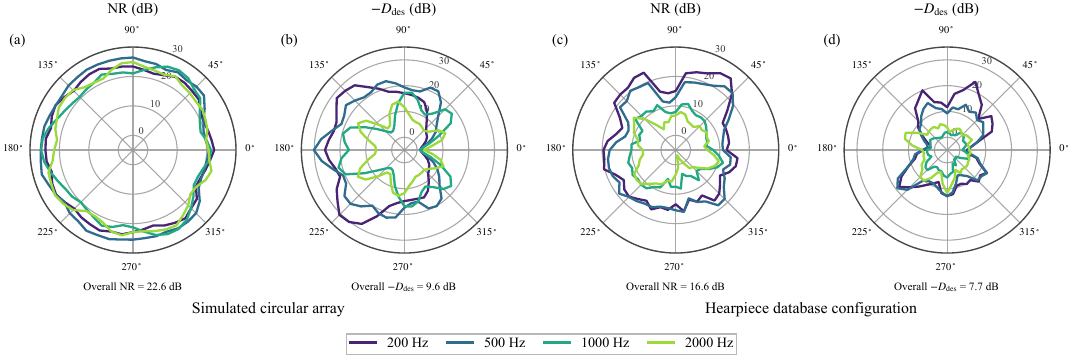}
\caption{Directional coverage of the $\lambda = 0.1$ DP-ANC models for the simulated circular array (left pair) and Hearpiece configuration (right pair). The desired direction is fixed at $0^\circ$, while the noise source sweeps 48 azimuths. Each ``Overall'' value averages four independent broadband evaluations per azimuth, while bandwise curves use a three-point circular moving average.}
\label{fig:directional_coverage}
\end{figure*}

The angular-resolution experiment in Fig.~\ref{fig:angular_resolution} retains the \textminus{}10~dB main-evaluation SNR and includes both DP-ANC and analytical SSANC at two trade-off settings. The full-azimuth coverage experiment in Fig.~\ref{fig:directional_coverage} uses a common \textminus{}5~dB pre-control SNR and 30~dB sensor-noise level for the simulated-array and Hearpiece configurations, and additionally reports frequency-resolved results in four analysis bands centered at 200, 500, 1000, and 2000~Hz. Each radial value corresponds to a control-filter bank estimated for that noise direction while the desired direction is fixed. Each broadband ``Overall'' value is the arithmetic mean of the per-case 20--2500~Hz energy ratios expressed in dB, evaluated over all 48 noise directions with four independent signals per direction.

\subsection{Measured Hearpiece-Path Validation}
\label{subsec:hearpiece_setup}

A separate experiment uses the Hearpiece database \cite{Denk2021Hearpiece} to evaluate the method with measured head-mounted primary transfer functions. We use Knowles Electronics Manikin for Acoustic Research (KEMAR) measurements at 48 horizontal azimuths and train a four-reference model for this microphone geometry.

The external references are the left and right concha and entrance microphones. The right-ear eardrum response defines the error position, and the right outer driver is the secondary source. Primary paths are used directly from the database. The secondary path is represented by a minimum-phase response with the measured magnitude. Signals are resampled to 8~kHz and processed over 20--2500~Hz using 256-tap control filters.

We train a separate Hearpiece model under a discrete measured-path protocol. Desired and noise directions are sampled from these measured azimuths with $\Delta\theta \geq 15^\circ$. Training SNR is sampled from $[-10,10]$~dB, and both desired and noise content are varied using Gaussian and UrbanSound8K signals with non-overlapping training and evaluation splits \cite{Salamon2014UrbanSound}. The Hearpiece model uses $\lambda = 0.1$. Independent sensor noise is added to each of the four reference channels and the error-microphone channel. Its level is set independently in each channel to yield a 30~dB sensor SNR relative to the local mixture. Filter estimation and evaluation follow the independent-signal protocol in Section~\ref{subsec:independent_realizations}. Directional evaluation fixes the desired direction at $0^\circ$, uses a pre-control SNR of \textminus{}5~dB, and sweeps the noise over all 48 measured azimuths. Each filter is evaluated on four independent 0.5~s signals.

\begin{figure}[!t]
\centering
\includegraphics[width=0.98\columnwidth]{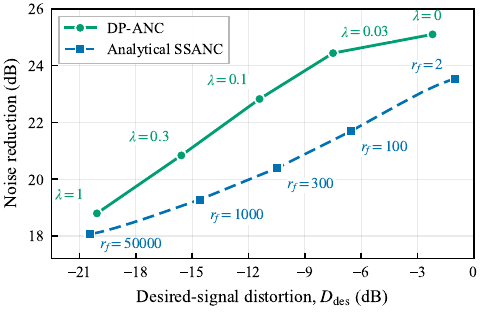}
\caption{Cancellation--preservation operating frontiers over the 3300 evaluation cases. Each marker corresponds to a separately trained or configured operating point.}
\label{fig:lambda_ablation}
\end{figure}

\section{Results and Discussion}
\label{sec:results}

\subsection{Average Cancellation and Desired-Signal Preservation}
\label{subsec:overall_results}

\begin{table}[!ht]
\centering
\caption{Average performance over 3300 evaluation cases under the main simulated-array configuration at a pre-control SNR of {\normalfont\textminus}10~dB. Values are mean $\pm$ standard deviation. The analytical SSANC row uses $r_{\beta} = r_f = 10^4$.}
\label{tab:multicontent_results}
\scriptsize
\setlength{\tabcolsep}{2.4pt}
\renewcommand{\arraystretch}{1.08}
\begin{tabular}{@{}lccc@{}}
\toprule
Method
& NR (dB) $\uparrow$
& $D_{\mathrm{des}}$ (dB) $\downarrow$
& $\mathrm{SNR}_{\mathrm{out}}$ (dB) $\uparrow$ \\
\midrule
ANC off
& $0.00 \pm 0.00$
& --
& $-10.00 \pm 0.01$ \\
Conventional Wiener
& $24.86 \pm 2.43$
& $-0.32 \pm 0.65$
& $-6.91 \pm 5.51$ \\
Beamformer hear-through
& $15.19 \pm 4.04$
& $1.35 \pm 1.10$
& $1.13 \pm 4.22$ \\
Analytical SSANC
& $18.28 \pm 3.64$
& $-19.14 \pm 4.12$
& $7.94 \pm 3.55$ \\
DP-ANC ($\lambda = 0.1$)
& $22.82 \pm 2.68$
& $-11.41 \pm 3.67$
& $12.52 \pm 3.36$ \\
\bottomrule
\end{tabular}
\end{table}

Figure~\ref{fig:lambda_ablation} shows that DP-ANC achieves higher mean noise reduction than analytical SSANC at matched mean desired-signal distortion across the five evaluated settings, with improvements ranging from 0.68 to 3.06~dB.

Table~\ref{tab:multicontent_results} complements the frontier comparison by reporting all three component metrics at representative fixed settings. The analytical SSANC setting is more preservation-oriented, with a mean $D_{\mathrm{des}}$ approximately 8~dB lower than DP-ANC, whereas DP-ANC provides higher NR and the highest output SNR among all methods.
 
As contextual endpoints, conventional Wiener ANC achieves the highest NR but its $D_{\mathrm{des}}$ near 0~dB indicates a desired-induced control response nearly as energetic as the desired component itself. Although the beamformer hear-through baseline achieves a desired-component correlation of 0.96 after temporal alignment, it introduces approximately 4.0~ms of delay, shifting the reproduced desired sound relative to its natural arrival time. The delayed reproduction may interfere with residual direct sound and introduce distortion.

Figure~\ref{fig:overall_performance} illustrates the desired and noise components before and after control for one example scene.

\subsection{Effect of the Preservation Weight $\lambda$}
\label{subsec:lambda_ablation}

Figure~\ref{fig:lambda_ablation} shows that increasing the preservation weight reduces desired-signal distortion at the expense of noise reduction. As the weight increases from 0 to 1, mean noise reduction decreases from 25.1 to 18.8~dB, while mean desired-signal distortion decreases from \textminus{}2.2 to \textminus{}20.1~dB. The weight therefore controls the balance between noise cancellation and desired-signal preservation across separately trained models.

\subsection{Angular Resolution and Directional Selectivity}
\label{subsec:directional_results}

Using the array-separability analysis in Section~\ref{subsec:main_simulated_configuration} as an interpretive reference, Fig.~\ref{fig:angular_resolution} compares DP-ANC and analytical SSANC as the desired--noise angular separation varies from $0^\circ$ to $90^\circ$. Two cancellation--preservation pairs are shown, comprising a moderate setting ($\lambda = 0.1$ versus $r_f = 1000$) and a preservation-oriented setting ($\lambda = 1$ versus $r_f = 10{,}000$). The parameters are selected so that each pair reaches similar $D_{\mathrm{des}}$ levels at large separations.

When the desired and noise directions coincide, their spatial responses are indistinguishable, so noise cancellation also affects the desired component. For the first pair of settings in Fig.~\ref{fig:angular_resolution}, increasing the angular separation from $0^\circ$ to $30^\circ$ improves both noise cancellation and desired-signal preservation: DP-ANC noise reduction increases from 21.5 to 23.3~dB, while its desired-signal distortion decreases from \textminus{}0.6 to \textminus{}9.4~dB. Analytical SSANC follows a similar trend, and both methods approach their respective operating levels at larger separations.

The full-azimuth experiment fixes the desired source and specified desired direction at $0^\circ$ and sweeps the noise source over 48 azimuths. Fig.~\ref{fig:directional_coverage} uses the $\lambda = 0.1$ DP-ANC model trained separately for each microphone configuration. Because every azimuth is evaluated with a control-filter bank estimated for that noise direction, the figure is a directional-coverage map rather than a fixed-filter beampattern.

For the simulated circular array, bandwise NR remains broadly high across azimuth, yielding 22.6~dB broadband overall NR. Preservation is more angle dependent. At the coincident $0^\circ$ direction, $-D_{\mathrm{des}}$ is only about 1--3~dB across the displayed bands, whereas it increases after the two source directions separate, consistent with Fig.~\ref{fig:angular_resolution}. The NR and $-D_{\mathrm{des}}$ panels therefore show where cancellation remains available and whether that cancellation is separated from the desired-direction control response.

\begin{figure}[!t]
\centering
\includegraphics[width=0.98\columnwidth]{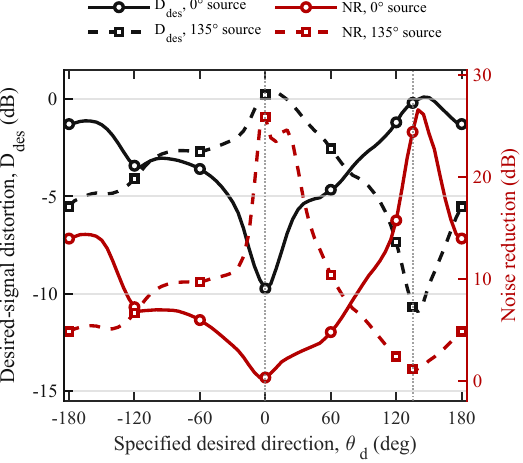}
\caption{Sensitivity to the specified desired direction, averaged over 32 fixed equal-power two-source scenes at $0^\circ$ and $135^\circ$. Within each scene, only the specified desired direction changes, while the acoustic scene and mixed observation remain fixed. Vertical dotted lines mark the physical source directions.}
\label{fig:command_sensitivity}
\end{figure}

Directional coverage alone does not show whether changing only the specified desired direction can exchange the roles assigned to sources in a fixed observation. Fig.~\ref{fig:command_sensitivity} therefore holds the acoustic scene and mixed-reference observation fixed and sweeps only the specified desired direction.

When the specified desired direction is aligned with either physical source direction, the aligned source achieves approximately \textminus{}10~dB $D_{\mathrm{des}}$ while the other source is attenuated by over 24~dB. The source-wise responses vary continuously as the specified direction sweeps between the two sources, demonstrating that the network redistributes cancellation and preservation according to the directional input.

\subsection{Validation on Measured Hearpiece Primary Paths}
\label{subsec:hearpiece_results}

The separately trained Hearpiece model maintains both noise cancellation and desired-signal preservation with a different microphone configuration and measured primary responses. As shown in Fig.~\ref{fig:directional_coverage}(c) and (d), it achieves 16.6~dB broadband noise reduction with a desired-signal distortion of \textminus{}7.7~dB. These results support the applicability of the proposed framework to measured acoustic paths and different microphone-array configurations.

\subsection{Computational Comparison}
\label{subsec:efficiency_results}

\begin{table}[!t]
\caption{Operation count required to obtain one control-filter bank.}
\label{tab:computational_requirements}
\centering
\footnotesize
\setlength{\tabcolsep}{4pt}
\renewcommand{\arraystretch}{1.08}
\begin{tabular*}{0.78\columnwidth}{@{\extracolsep{\fill}}cc@{}}
\toprule
Method & Operation count (GMACs) \\
\midrule
DP-ANC & $0.00946$ \\
Analytical SSANC & $\approx29.4$ \\
\bottomrule
\end{tabular*}
\end{table}

Table~\ref{tab:computational_requirements} compares the operation counts required to obtain one control-filter bank. DP-ANC requires 0.00946~GMACs, compared with approximately 29.4~GMACs for analytical SSANC. Both counts exclude the subsequent sample-by-sample FIR filtering.

\subsection{Limitations}
\label{subsec:limitations}

The evaluation focuses on static scenes containing one desired source and one noise source. Extending the component-separated objective to multiple simultaneous sources is a natural next step. The main configuration is anechoic. The Hearpiece validation introduces measured primary responses. Reverberation, moving sources, path mismatch, and closed-loop hardware implementation remain for future work. The network currently operates on a 0.5~s observation block and is trained for a specific microphone geometry, motivating streaming updates and cross-device adaptation. Its near-direction behavior remains governed by the spatial separability of the reference array, as characterized in Fig.~\ref{fig:angular_resolution}.

\section{Conclusion}
This work developed a component-separated soft objective and a direction-conditioned control-filter estimation network for direction-preserving ANC. The network estimates the complete control-filter bank in a single forward pass while retaining the conventional feedforward ANC signal path. In the simulated evaluation, the proposed method achieved higher mean noise reduction than analytical SSANC at matched mean desired-signal distortion. The angular-resolution results show how reference-array spatial separability shapes the cancellation--preservation trade-off. Validation with a separately trained Hearpiece model further demonstrates that the same architecture can balance noise cancellation and desired-signal preservation under measured primary-path conditions.

Future work will extend the component-separated objective to multi-source and time-varying scenes and develop streaming updates and cross-device adaptation for continuously changing acoustic environments.
\section*{Acknowledgment}
The authors would like to thank Dr. Tong Xiao for sharing the source code associated with his prior work on spatially selective active noise control.

\bibliographystyle{IEEEtran}
\bibliography{references}

\end{document}